\documentclass[aps, pra, showpacs, superscriptaddress,10pt]{revtex4-1}
\usepackage{graphicx}
\usepackage{dcolumn}
\usepackage{bm}
\usepackage{amsmath}
\usepackage{euscript}
\usepackage{mhchem}
\usepackage{gensymb}
\usepackage[utopia]{mathdesign}

\begin{document}

\title{Critical behaviour in the optimal generation of multipartite entanglement}

\author{M. G. M. Moreno and Fernando Parisio}
\affiliation{Departamento de F\'{i}sica, Universidade Federal de Pernambuco, 50670-901, Recife, Pernambuco, Brazil}


\begin{abstract}
Two systems whose correlations cannot be classically accounted for display the simplest instance of quantum entanglement. 
Although this two-party association has caused a revolution in the foundations and uses of quantum mechanics, 
genuine entanglement among several parties unveils a whole new class of
phenomena and applications. In this work we suggest a way to prepare Dicke states from a tunable source of bipartite entanglement to investigate foundational issues. 
The scheme has the following distinctive features:  (i) it relies on controlled information loss and {\it unentangled} measurements; (ii) irrespective of the source entanglement, whenever a Dicke state is produced, it is ideal; (iii) the optimal entanglement of the bipartite source undergoes a second-order-like transition depending on the parameters of the Dicke state to be produced. These properties lead to asymptotic results on the entanglement between any qubit belonging to a Dicke state and the remaining qubits.
\end{abstract}

\flushbottom
\maketitle
%
%
\thispagestyle{empty}

\section*{Introduction}
Quantum entanglement involving two parties has led to groundbreaking advances, of which preeminent examples are Bell inequalities and nonlocality \cite{bell2004speakable}, teleportation \cite{bennett1993teleporting}, and dense coding \cite{bennett1992communication,mattle1996dense}. It would be natural to think, at a first sight, that multipartite entangled systems would present the same features in a larger scale. However, since the seminal study on the nonlocality of certain tripartite states in the early 90's  \cite{greenberger1990bell}, it became gradually clear that completely new phenomena and potential applications could arise. It is now known that $n$ particles can be highly entangled without any pairwise correlation: entanglement appears in many different, inequivalent forms, of which the Einstein-Podolsky-Rosen (EPR) type is the simplest instance  \cite{li2012classification,dur2000three}. As for applications, there are, e. g., paradigms for quantum computation which rely on the feasibility of cluster states of several qubits \cite{raussendorf2001one}. 

It is, then, clear that conceiving means to produce multipartite entangled states is of relevance. To date, there are three ways to meet this goal: controlled interactions between qubits, measurements in entangled basis, and indistinguishability of identical particles. The first way relies on the fact that initially uncorrelated interacting subsystems may become entangled. Of course, the interactions must be finely tuned and the qubits protected from noise. The second possibility is related to the fact that, to any orthogonal entangled basis corresponds a physical observable that can be measured, in principle, leaving the system entangled.  There are, however, serious provisos regarding this simplistic picture. Firstly, measurements are commonly destructive, photo-detection for instance, so, after their realisation there remains no system whatsoever. This problem can be circumvented if bipartite entanglement is an available resource. So, instead of using $n$ qubits, one employs $n$ EPR pairs and proceed with the $n$-partite measurement on one particle of each pair. The remaining qubits will end up, with some non-zero probability, in a $n$-partite entangled state, see, e. g.,  \cite{moreno2016remote}. This brings the second difficulty. In practice, it is very hard to make full measurements in entangled bases, even in the bipartite case \cite{vaidman1999methods,lutkenhaus1999bell}. The third way is more related to a fundamental principle than to deliberate procedures, see however \cite{bose}. It simply amounts to the fact that two indistinguishable electrons, e. g., can only exist in an entangled state.
 
In this article we present a procedure for creating arbitrary Dicke states, $|D^{(k)}_n\rangle$, of $n$ particles and $1\le k \le n-1$ excitations. It employs a combination of induced indistinguishability and unentangled Fock measurements. The protocol has the property that, irrespective of how poor is the source of entanglement, every time a Dicke state is created, it is ideal. In addition, the probability of creating $|D^{(k)}_n\rangle$ behaves similarly to a thermodynamic potential during a second-order phase transition, as $n$ grows, where the entanglement of the optimal source undergoes a qualitative change at a critical $n$. Finally, we establish exact results on the asymptotic entanglement between any qubit belonging to $|D^{(k)}_n\rangle$ and the rest of the system.
\section*{Entanglement lifting}
We begin by sketching an optical realisation of the scheme in the simple case of three photon pairs. It illustrates what we refer to as entanglement {\it lifting}. Differently from usual entanglement concentration, we start with $M$ bipartite entangled pairs and, in the end, we obtain $j$ $n$-partite entangled systems. More precisely, in the pure case, we will consider the process $|\phi\rangle^{\otimes M} \longrightarrow |\Psi\rangle^{\otimes j}$ with $|\phi\rangle \in {\cal H}_{i}$ and  $|\Psi\rangle \in {\cal H}_{f}$, where not only $j<M$, but also $D>d$, with $d={\rm dim}{\cal H}_{i}$, and $ D={\rm dim}{\cal H}_{f}$. It will become clear that lifting also entails concentration. 

Consider three identical sources of pairs of polarisation-entangled photons (alternatively one can consider that there is a single source producing the pairs which are posteriorly distributed). We initially assume that each pair is maximally entangled, with state $|\phi^+\rangle=(|00\rangle+|11\rangle)/\sqrt{2}$, with 0 (1) denoting horizontal (vertical) polarisation. One photon of each pair is delivered to Alice, while the others are sent to Bob, Brian and Brandon. After this, the total quantum state reads $|\Psi_0\rangle=|\phi^+\rangle^{\otimes 3}=2^{-3/2}\sum_{j,k,l}|ijk\rangle_A\otimes |ijk\rangle_B$,
where $A$ stands for Alice and $B$ for the spatially separated system of Brandon, Brian and Bob (fig. \ref{fig1}). 

Alice synchronizes her photons, e. g., using quantum memories \cite{felinto2006conditional,yuan2007synchronized,makino2016synchronization} (process denoted by S) before sending them to a polarization beam splitter (PBS). It is important that the three wave packets overlap so that the photons are undistinguishable when they arrive at the PBS. With this, Alice intentionally discards the information on the former holder of each photon.

After passing the PBS, the photons proceed to detectors capable of discriminating the Fock state in each spatial mode of Alice's system \cite{divochiy2008superconducting,0953-2048-28-10-104001}. The indistinguishability induces a natural partition of Alice system's Hilbert space into subspaces generated by kets with a fixed number of photons with a given polarisation, so that, states like $|001\rangle$ and $|010\rangle$ coalesce into $|2~ {\rm horizontal};1 ~{\rm vertical}\rangle$. In this way, we have four detection possibilities: $|000\rangle  \rightarrow  |3;0\rangle$, $\{ |001\rangle , |010\rangle , |100\rangle\}\rightarrow  |2;1\rangle$, $\{|011\rangle , |101\rangle , |110\rangle\} \rightarrow  |1;2\rangle$, and $|111\rangle  \rightarrow |0;3\rangle$.
The first entry in the kets on the right-hand side gives the photon number of path $2$, and the second entry the photon number of path $1$, see fig. \ref{fig1}. 
It is clear that this process is non unitary since it can map orthogonal states into the same final state. This is due to the fact that synchronisation can only be achieved through the interaction between the photons and some ancillary systems which, generally speaking, store the information on the initial delay between the photons. In \cite{yuan2007synchronized}, where the possibility of using this technique in the scalable generation of photonic entanglement is already mentioned, the auxiliary systems consist of Rubidium atom ensembles. The global unitary evolution of the larger system leads to the non unitary evolution of the reduced system of the photons to be synchronised.

In the full indistinguishability situation, just before the detection, the state is proportional to:
\begin{eqnarray}
\nonumber
|3;0\rangle |000\rangle +\sqrt{3}|2;1\rangle\left(\frac{1}{\sqrt{3}}|001\rangle +\frac{1}{\sqrt{3}}|010\rangle+ \frac{1}{\sqrt{3}}|100\rangle\right) 
\nonumber
+ \sqrt{3}|1;2\rangle \left(\frac{1}{\sqrt{3}}|011\rangle + \frac{1}{\sqrt{3}}|101\rangle+\frac{1}{\sqrt{3}}|110\rangle \right) + |0;3\rangle|111\rangle.
\end{eqnarray}
Therefore, after the detection of Alice's photons there are four possible states left to the B-group. If all three photons are either detected in path 1 or in path 2 (probability 1/8 for each event), then, B-group's state is separable and, after appropriate classical communication, it is disposed. With probability $3/8$ two photons are detected in path 2 and one in path 1 and, then, B-group's state is already a $W$ state, $|D^{(1)}_3\rangle\equiv|W_3\rangle \propto |001\rangle +|010\rangle+|100\rangle$. Lastly, also with probability 3/8, two photons are detected in path 1 and one in path 2. In this case, Alice communicates the B-group members that each of them have to perform a bit-flip operation in his qubit, and, in the end, ideal $W$ states are prepared with probability $3/4$. 
\section*{Results}
We now generalise the lifting procedure to the case where $n$ EPR pairs are used per run. We will see that this leads to the production of arbitrary, unambiguously heralded Dicke states. The initial state is given by $|\Psi_0\rangle =|\phi^+ \rangle^{\otimes n}$, and it is easy (but crucial) to see that it can be written as $|\Psi_0\rangle 
=\left(\frac{1}{2}\right)^{n/2}\sum|j_1 j_2 ...j_n\rangle_A \otimes |j_1 j_2 ...j_n\rangle_B$, where $j_i=0,1$.
Again, one photon of each pair is sent to Alice and the others to each of the elements of the B-group, now composed by $n$ parties. Alice's photons are synchronised and sent to the PBS. Given the indistinguishability, besides $|00...000\rangle_A  \rightarrow  |n;0\rangle$ and $|11...111\rangle_A  \rightarrow  |0;n\rangle$, we have the following correspondences: $\{ \hat{\cal P}_j^{\left(1\right)}|00...001\rangle_A\}_j  \rightarrow  |n-1;1\rangle$, $\cdots$, $\{\hat{\cal P}_j^{\left(n-1\right)}|01...111\rangle_A\}_j  \rightarrow  |1;n-1\rangle$, where $\hat{\cal P}_j^{\left(k\right)}$ represents the $j$th non-trivial permutation of the ket entries. Thus, for each $k$, we have $j=1,2,..., \left.n \choose k\right.$. It is simple to show that the global state of the system immediately before the detection of Alice's photons is
\begin{eqnarray}
\label{Dicke}
|\Psi\rangle  =  \left(\frac{1}{2}\right)^{\frac{n}{2}} \Bigg[ |n;0\rangle_A|00...00\rangle_B + |0;n\rangle_A |1...111\rangle_B 
 + \sum_{k=1}^{n-1}\left. n \choose k\right. ^{\frac{1}{2}}|n-k;k\rangle_A|D^{(k)}_n\rangle_B\Bigg],
\end{eqnarray}
where
\begin{eqnarray}
|D^{(k)}_n\rangle \equiv \left. n \choose k\right.^{-\frac{1}{2}}\sum_{j=1}^{n\choose k}\hat{\cal P}_j^{\left(k\right)}|\underbrace{0...0}_{n-k}\overbrace{1...11}^k\rangle
\end{eqnarray}
is a $n$-partite Dicke state with $1 \le k \le n-1$ excitations. Note in Eq. ({\ref{Dicke}) the perfect correlation between Alice's output of an unentangled Fock measurement and the production of a specific ideal Dicke state shared by the elements of the B-group. 
The probability of having output $|n-k;k\rangle$, or yet, to remotely produce $|D^{(k)}_n\rangle$, is $\left.n \choose k\right.2^{-n}$. Dicke states with different number of excitations are generally inequivalent under local operations and classical communications (LOCC). The obvious exception occurs for $k$ and $n-k$ excitations, since
$|D^{(k)}_n\rangle=\hat{\sigma}_x^{\otimes n}|D^{\left(n-k\right)}_n\rangle$.
Therefore, after Alice communicates her outcome, the B-group may transform all the states with $2k>n+1$ ($2k>n$), into states with $2k\le n-1$ ($2k< n$), for $n$ odd (even). We will assume that this procedure is always adopted in the remainder of this manuscript. 
For $n$ even and $k=n/2$, we already have $|D^{(k)}_n\rangle=|D^{\left(n-k\right)}_n\rangle$.

We are in a position to further extend the scheme to the case of tunable sources producing pairs with arbitrary entanglement  \cite{PhysRevLett.83.3103,xu2015experimental,mizutani2015robustness,mendes2015femtosecond}: $|\phi\rangle = a|0_A 0_B\rangle + b|1_A 1_B \rangle$. 
In this case the total initial state $|\Psi_0\rangle$ can be written as:
\begin{eqnarray}
\label{GEN}
\nonumber
\sum_{j_1,\space j_2,\space...,\space j_n=0,1}a^n\left(\frac{b}{a}\right)^{\sum_{\ell=1}^nj_{\ell}}|j_1 j_2 ...j_n\rangle_A \otimes |j_1 j_2 ...j_n\rangle_B.
\end{eqnarray}
Under full indistinguishability, the surprising result is that after the PBS, and before detection, the state reads:
\begin{eqnarray}
\label{psitot}
|\Psi\rangle = a^n|n;0\rangle_A|00...00\rangle_B + b^n|0;n\rangle_A |1...111\rangle_B
+ \sum_{k=1}^{n-1}a^{n-k}b^k \left. n \choose k\right. ^{\frac{1}{2}}|n-k;k\rangle_A|D^{(k)}_n\rangle_B,
\end{eqnarray}
that is, despite the asymmetry of the source states, whenever an entangled state is produced it is still a perfect $|D^{(k)}_n\rangle$. It is of great importance to note that, had we kept all initial information, namely, the distinguishability of Alice's photons, then entangled measurements would be required to leave the B-group with Dicke states.
The unbalanced character of the bipartite source, rather than affecting the ideality of the outputs, only changes their probabilities of occurrence, which are given by
\begin{eqnarray}
\label{prob1}
{P}^{(k)}_n=\left. n\choose k \right.\left(\left|a^{n-k}b^k\right|^2+\left|a^kb^{n-k}\right|^2\right)\, {\cal G}_{n,k},
\end{eqnarray}
with ${\cal G}_{n,k}=1$ for $n$ odd and  ${\cal G}_{n,k}=(1+\delta_{k,n/2})^{-1}$ for $n$ even, and, with the LOCC-equivalence between $k$ and $n-k$ excitations already considered. 

In fig. \ref{fig2} we display (a) ${P}^{(1)}_n$ and (b) ${P}^{(2)}_n$, as functions of $|a|^2$, for selected values of $n$. The left panel shows that for $n=3$, the optimal source corresponds to maximally entangled states ($|a|^2=1/2$). This is also true for $n=4$, but, in this case there is a broad plateau around $|a|^2=1/2$. For $n\ge 5$,  ${P}^{(1)}_n$ is maximal for non-maximally entangled sources and, two symmetric optimal values of $|a|^2$ appear.  By using the general expression (\ref{prob1}) one can analytically determine the bifurcation condition:
\begin{equation}
\label{ineqn}
n> 2k+\frac{1}{2}+\sqrt{2k+\frac{1}{4}}\equiv \eta_c.
\end{equation}
Since the right-hand side of the above equation is not necessarily integer, we define $n_c=\lceil \eta_c \rceil$, which stands for the ceiling function (smallest following integer). For $k=1$ and $k=3$, $\eta_c=n_c=4$ and $\eta_c=n_c=9$, respectively. In these cases the critical value is indeed integer and one can see the typical behaviour shown in fig. \ref{fig2}(a) for $n=4$. In contrast, for $k=2$, $\eta_c \approx 6.56$ ($\Rightarrow n_c=7$), see fig. \ref{fig2}(b), and the exact critical point is passed by, for, there is no $n$ for which the plateau appears: for $n=6$ there is a single maximum in ${P}^{(2)}_n$, whereas for $n=7$ there are two maxima. However, the properties hereafter derived are the same no matter if $\eta_c$ is an integer or not.
In fig. \ref{fig3}(a) we plot the optimal source parameter $|a_{opt}|^2\equiv |\tilde{a}|^2$ against $n$ for $k=1$ ($n_c=4$) and $k=3$ ($n_c=9$). The optimal source state remains maximally entangled up to $n=n_c$, after which a bifurcation develops. 

These features remind us of the classical Landau theory \cite{callen2006thermodynamics,pathria1972statistical} of second order phase transitions, with the role of the thermodynamic potential being played by  ${P}^{(k)}_n$, $|a|^2$ as the order parameter, and $n$ inversely related to the temperature. Given this similarity, we set to find further evidence to support our analogy. Figure \ref{fig3}(b) displays the  probability ${P}^{(3)}_n$ as $n$ varies. The filled bullets represent ${P}^{(3)}_n$ if one employs a maximally entangled source, $|a|^2=1/2 ~ \forall ~n$, leading to a steep decay toward zero. As soon as $n>n_c$, if, instead, one uses the optimal, but less entangled source determined by the lateral maxima, the decay follows a power law with a {\it finite}, non-vanishing asymptotic value. This regime is represented by the stars in fig. \ref{fig3}(b). We found that the state of the optimal bipartite source for the production of $|D^{(k)}_n\rangle$, is asymptotically given by:
\begin{equation}
\label{opt}
|\tilde{\phi}_{n}^{(k)}\rangle \rightarrow \sqrt{\frac{k}{n}}\,|00\rangle+\sqrt{1-\frac{k}{n}}\,|11\rangle, \;n\rightarrow \infty,
\end{equation}
with $k$ finite. The other optimal state is obtained via $|a| \rightleftarrows |b|$ [Lower branches in fig. \ref{fig3} (a) tend to Eq. (\ref{opt})]. Thus, the optimal source tends to a collection of quasi-separable states. Yet, by using it we end up with a non-zero probability of obtaining $n$-partite entangled states. In this limit, the behaviour of the optimal probability, $\tilde{P}^{(k)}_n$, is given by:
\begin{equation}
\label{asympa}
\tilde{P}^{(k)}_n = \tilde{P}^{(k)}_{\infty}\left[1+k\, (2 n)^{-1}\right] +O\left(n^{-2}\right),
\end{equation}
where the constant asymptotic probability reads 
\begin{equation}
\label{asympb}
\lim_{n\rightarrow \infty}\tilde{P}^{(k)}_{n}\equiv  \tilde{P}^{(k)}_{\infty}=\frac{k^{k}\,e^{-k}}{k!}.
\end{equation}
Although these are asymptotic results, the convergence is fast for small $k$'s .  The stars in fig. \ref{fig3} (b) are close to the continuous curve [$0.224(1+3/2 n)$] already for $n>15$, see the lower-order non-constant term in the asymptotic expression for $\tilde{P}^{(k)}_n$, derived in the methods section.

The properties derived so far provide information on the entanglement of Dicke states. Result (\ref{asympb}) leads to the conclusion that, if entangled pairs are an available resource, the probability to create Dicke states as $n\rightarrow \infty$, with a finite number of excitations $k$ is {\it non-vanishing even if the source has an arbitrarily low entanglement per pair}. For $k=1$, e. g., one obtains $\tilde{P}^{(1)}_{\infty}\approx 0.368$, with $|\tilde{a}|^2\rightarrow 1/n$. This is the signature of entanglement concentration: the exchange of a large number of copies with low entanglement by a small quantity of more entangled systems  \cite{bennett1996concentrating}. Here, in addition to concentration, the initial bipartite entanglement is lifted to a larger Hilbert space.
	
Finally, we derive asymptotic results on the amount of entanglement between any single qubit which is part of a system in a Dicke state and the rest of the system. Therefore we address the partition $($Alice, $ B_1,\cdots, B_{j-1},B_{j+1},\cdots, B_N|B_j) \equiv (I|II) $, where $B_j$ represents any of the elements in the B group (due to the symmetry of the Dicke state the analysis does not depend on which qubit is being singled out). Note that, after Alice's local measurements this same partition reads $(B_1,\cdots, B_{j-1},B_{j+1},\cdots, B_N|B_j)$, which indeed refers to the entanglement between an arbitrary B-group member and the rest of the group. Suppose one intends to produce $|D^{(k)}_n\rangle$ with $k$ finite and $n$ arbitrarily large from a reservoir of optimal bipartite states, given by  Eq. (\ref{opt}).  The initial entanglement between $I$ and $II$ is precisely the entanglement between $B_j$ and the rest of the system, i. e., $E(|\tilde{\phi}^{(k)}_n\rangle)$, once local operations and classical communications should not increase the entanglement between any bipartition, the final amount of entanglement between theses parts are upperbounded by $E(|\tilde{\phi}^{(k)}_n\rangle)$, where $E$ is an arbitrary bipartite measure or monotone. Nevertheless it must also be lowerbounded by the amount of entanglement $\tilde{P}^{(k)}_n\times E_{(I|II)}(|D^{(k)}_n\rangle)$ between $B_j$ and the rest of the B group --at this point Alice's part is completely factorable and plays no role. This holds because we are disregarding the contributions coming from the other Dick states that may be created. So, we must have, 
\begin{eqnarray}
\label{constr}
E(|\tilde{\phi}^{(k)}_n\rangle) > \tilde{P}^{(k)}_nE_{(I|II)}(|D^{(k)}_n\rangle).
\end{eqnarray}
As $n \rightarrow \infty$ the left-hand side goes to zero from above, the same must be true for the positive-definite quantity in the right-hand side. However, in this limit the asymptotic probability is non-vanishing, which demands
\begin{eqnarray}
E_{(I|II)}(|D^{(k)}_n\rangle)\rightarrow 0, \;\;n\rightarrow \infty.
\end{eqnarray}
This result is to be contrasted with the entanglement referring to the same partition of a system in a GHZ state of $n$ qubits: $|GHZ_n\rangle=2^{-1/2}(|0\rangle^{\otimes n}+|1\rangle^{\otimes n})$, which gives $E_{(I|II)}(|GHZ_n\rangle)=1$ ebit, for arbitrary $n$.
As a specific example let us consider the 2-tangle as the bipartite measure. In this case, equation (\ref{constr}) reads, 
${\tau_2}_{(I|II)}(|D^{(k)}_n\rangle) <(4k/\tilde{P}^{(k)}_n)\, n^{-1}+O(n^{-2})$. Therefore, the 2-tangle between an arbitrary qubit in a Dicke state and the rest of the system must go to zero, at least, as fast as $n^{-1}$, as $n\rightarrow \infty$ with $k$ fixed.

\section*{Discussion}

We presented an efficient protocol to create Dicke states without the need of entangled measurements, a major technical difficulty. The efficiency is due to a high success probability and to the fact that whenever a Dicke state is produced it is ideal.
In optimally executing this protocol the tunable source states present a behaviour that is analogous to a second-order phase transition in thermodynamics. In particular, the optimal source is not necessarily made of maximally entangled pairs.

Although our conclusions rely on the fact that our scheme is possible, in principle, an experiment seems to be already feasible for a relatively small number of parties. The most important ingredients, namely, tunable bipartite sources \cite{PhysRevLett.83.3103,xu2015experimental,mizutani2015robustness,mendes2015femtosecond}, photon-number-resolving detectors \cite{divochiy2008superconducting,0953-2048-28-10-104001} and synchronisation techniques \cite{felinto2006conditional,yuan2007synchronized,makino2016synchronization} are presently available, although, the conjunction of these elements may pose technical difficulties. The result of such an implementation would be the production of highly entangled multipartite states with a strong robustness against poor sources and with an increasing success rate for large $n$'s.  We finally call attention to the fundamental relation between the information loss induced by synchronisation and the simpler nature of the required measurements. This counterintuitive fact has been first reported, in a quite distinct context, in \cite{fortescue2007random} and it deserves further investigation.

\section*{Methods}

{\small
Here we give the main steps to demonstrate  equations (\ref{opt}) to (\ref{asympb}).
We intend to find the optimal probability of obtaining the state $|D_n^{\left(k\right)}\rangle$ in the situation $n>n_c$. For $n\ne k/2$ the probability reads
\begin{equation}
\label{A1}
P_n^{\left(k\right)}=\left. n \choose k\right.\left[A^{\left(n-k\right)}\left(1-A\right)^k + A^k\left(1-A\right)^{n-k}\right],
\end{equation}
where $A=\left|a\right|^2$. Initially we had numeric evidence that the optimized probability would follow a power law to reach a finite value as $n \rightarrow \infty$.
We, therefore, set to look for a source state that would support this behaviour.
Looking at Eq. (\ref{A1}), since $0\le A \le 1$, we see that as $n\rightarrow\infty$, we must have either $A\rightarrow0$ [lower branches in fig. 3 (a)] or $A\rightarrow 1$ [uper branches in fig. 3 (a)] in order to observe a $P_n^{\left(k\right)}$ finite. These two choices are equivalent due to the symmetry of equation (\ref{A1}), so we address the lower branch  $A\rightarrow 0$ as $n\rightarrow\infty$. In this case, the first therm in Eq. (\ref{A1}) vanishes, leading to the following result:
\begin{equation}
\nonumber
\lim_{\substack{n \to \infty \\ A \to 0}}P_n^{\left(k\right)}=\frac{n\left(n-1\right)...\left(n-k+1\right)}{k!}A^k\frac{\left(1-A\right)^n}{\left(1-A\right)^k}
\end{equation}
The product $n\left(n-1\right)\cdot...\cdot\left(n-k+1\right)$ gives rise to a term of order $n^k$ plus lower-order terms. So the only way to keep $P_n^{\left(k\right)}$ finite is to assume that $A \sim n^{-1}$, say, $A=\frac{\alpha}{n}$. With this we indeed obtain the finite asymptotic value:
\begin{equation}
\nonumber
\lim_{\substack{n \to \infty \\ A \to 0}}P_n^{\left(k\right)} = \lim_{\substack{n \to \infty}}\frac{\alpha^k}{k!}\left(1-\frac{\alpha}{n}\right)^n=\frac{\alpha^k}{k!}e^{-\alpha}\equiv P^{(k)}_{\infty},
\end{equation}
with $\alpha$ to be determined. Since we are seeking the optimal source, we simply maximize $P_{\infty}^{\left(k\right)}$:
\begin{equation}
\nonumber
\frac{d}{d\alpha}P_{\infty}^{\left(k\right)}=0 \Rightarrow \alpha=k,\;A_{opt}\equiv \tilde{A}=\frac{k}{n},
\end{equation}
which justifies Eqs. (\ref{opt}) to (\ref{asympb}). Equation (\ref{asympa}) is obtained by collecting the following lower order term, leading to
\begin{equation}
\nonumber
\tilde{P}_n^{\left(k\right)}\sim \frac{k^k}{k!}e^{-k}\left(1+\frac{k}{2n}\right).
\end{equation}
In general, for small $k$ this regime is quickly reached and, as the number of excitations increases (remaining finite), we observe a longer transient.

FIGURE CAPTIONS\\

FIG1: Three photon pairs are produced in an EPR polarisation state. One photon of each pair is sent to Alice, while the B-group members receive one photon each. Alice send her synchronised photons to a PBS, after which they are detected, leaving the B-group with an ideal Dick state with probability $p=0.75$.\\

FIG2: Probability of producing Dicke states with (a) one excitation and (b) two excitations for selected values of $n$. As $n$ grows there is a transition from functions with a single maximum at $|a|^2=1/2$ to curves with two maxima.\\

FIG3: (a) ``Pitchfork'' bifurcation diagram of $|a_{opt}|^2\equiv |\tilde{a}|^2$ versus $n$ for $k=1$ (diamonds) and $k=3$ (bullets). (b) Probabilities of getting creating a Dicke state with 3 excitations versus $n$. If EPR pairs are used $P^{(3)}_n$ drops exponentially (bullets). If the optimal source is used, it undergoes a power-law decay (stars).\\

\section*{Acknowledgements}

This work received financial support from the Brazilian agencies Coordena\c{c}\~ao de Aperfei\c{c}oamento de Pessoal de N\'{\i}vel Superior (CAPES), Funda\c{c}\~ao de Amparo \`a Ci\^encia e Tecnologia do Estado de Pernambuco (FACEPE), and Conselho Nacional de Desenvolvimento Cient\'{\i}fico e Tecnol\'ogico (CNPq). 

\section*{Author contributions statement}

M. G. M. M. and F. P devised the protocol and derived the results. F. P. wrote the paper and M. G. M. M. prepared the figures.

\section*{Additional information}

\textbf{Competing financial interests} The authors declare no competing financial interests.\\

\newpage

\begin{figure}
\centering
\includegraphics[scale=0.3,angle=0]{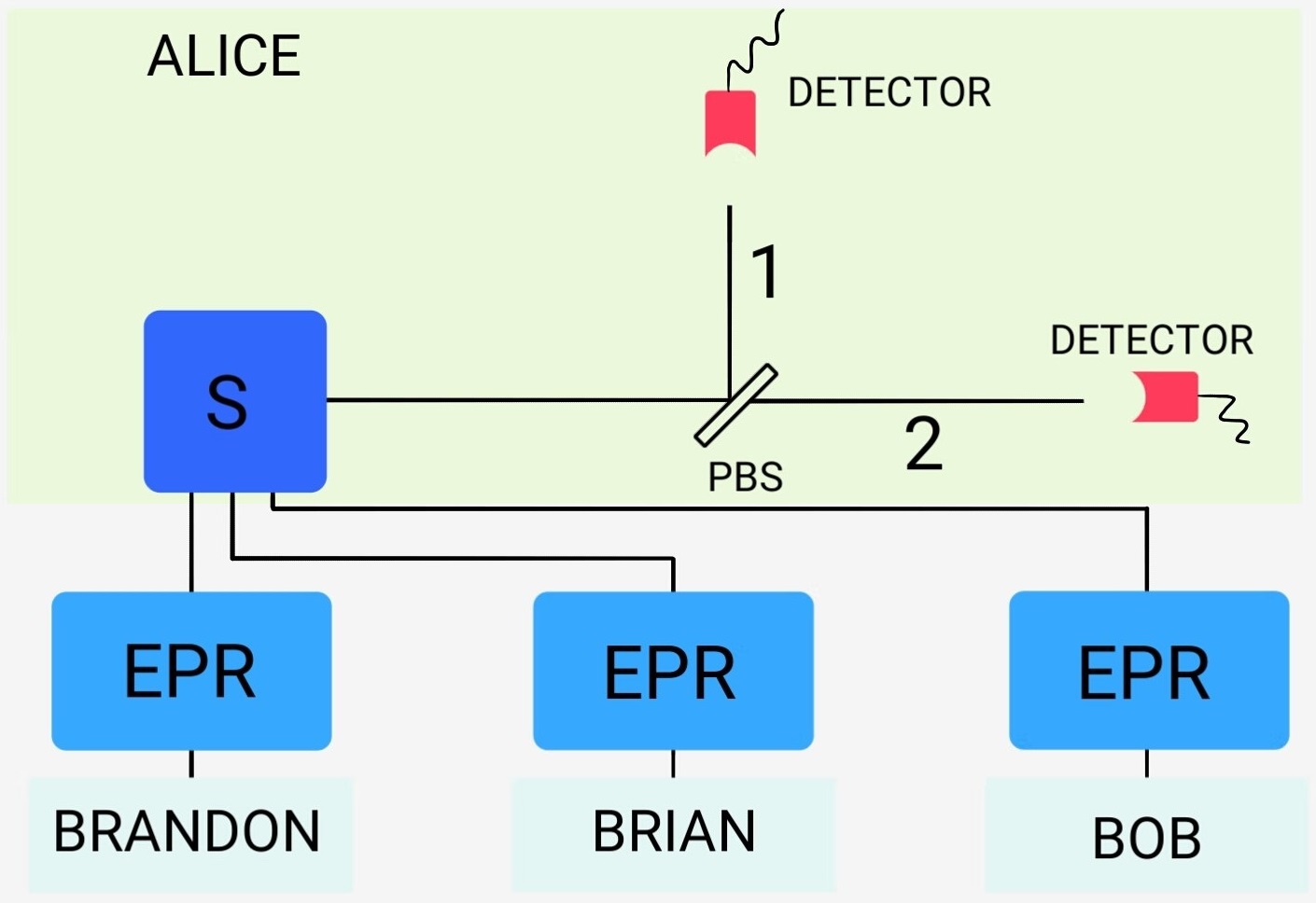}	
\caption{}
\label{fig1}
\end{figure}
\begin{figure}
\centering
\includegraphics[scale=0.3]{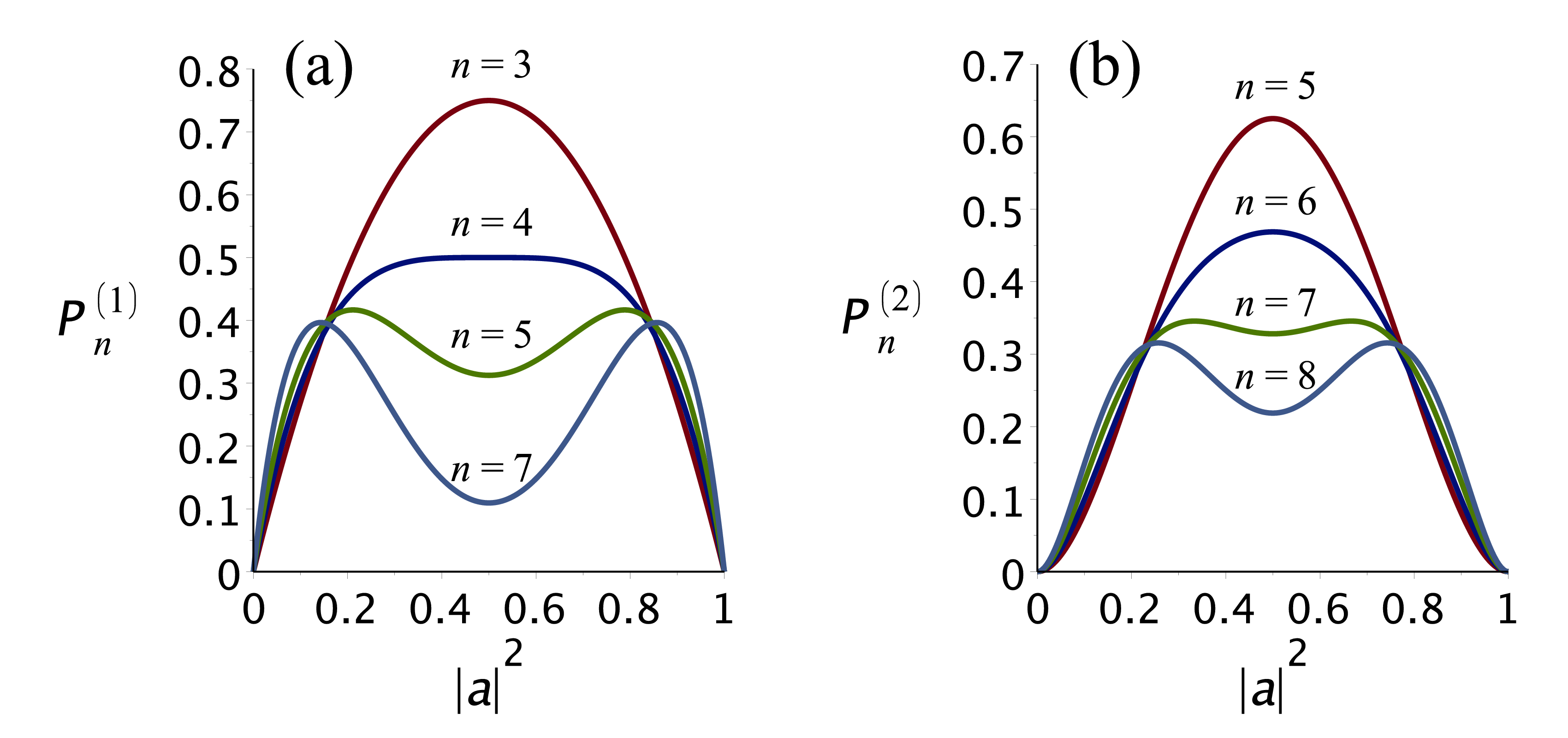}
\caption{}
\label{fig2}
\end{figure}
\begin{figure}
\centering
\includegraphics[scale=0.3]{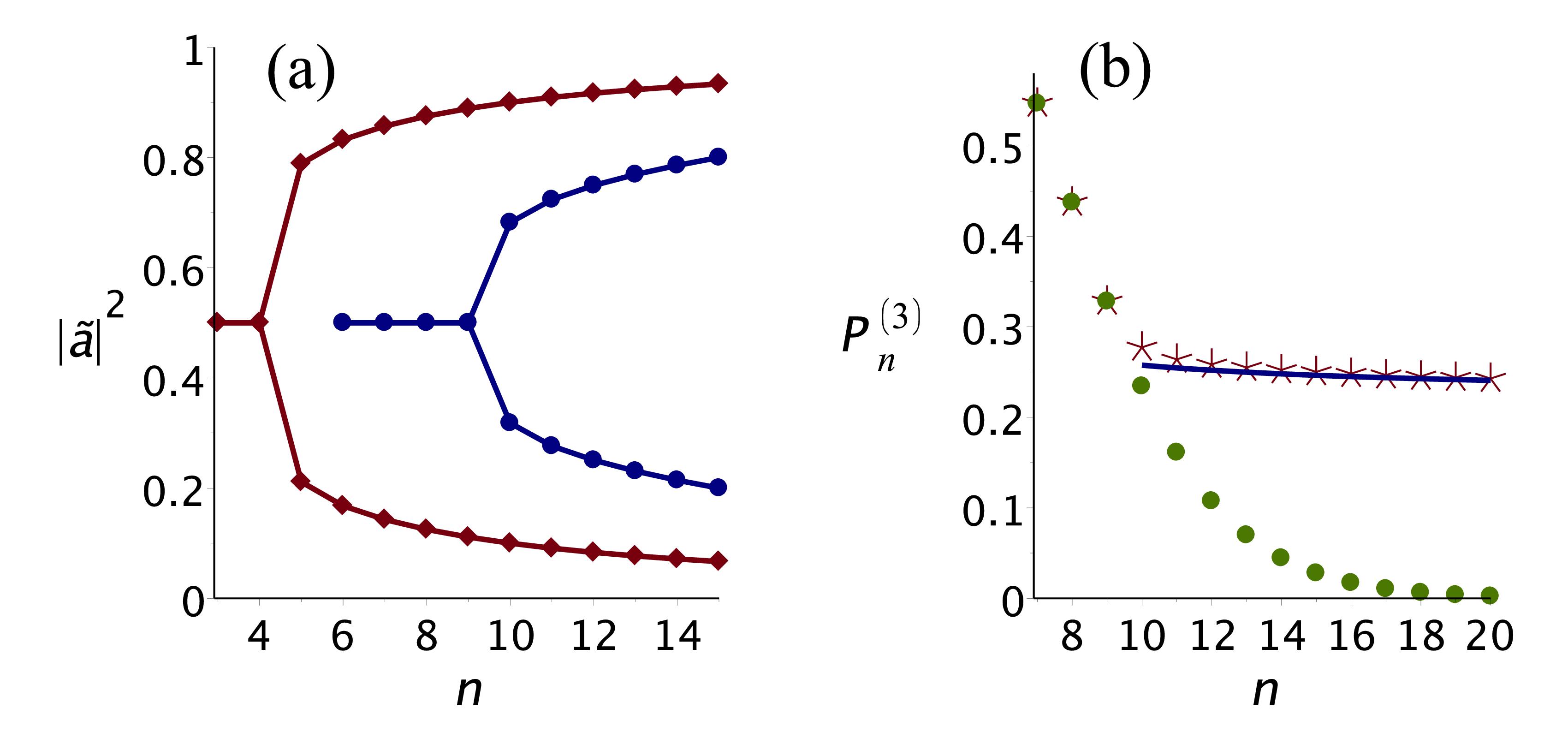}
\caption{}
\label{fig3}
\end{figure}


\begin{thebibliography}{99}
\bibitem{bell2004speakable} Bell, J. S., {\em Speakable and unspeakable in quantum mechanics: Collected papers on quantum philosophy}  (Cambridge university press, 2004).

\bibitem{bennett1993teleporting} Bennett, C. H., \textit{et al.}, Teleporting an unknown quantum state via dual classical and Einstein-Podolsky-Rosen channels. \textit{Phys. Rev. Lett.} \textbf{70}, 1895 (1993).

\bibitem{bennett1992communication} Bennett, C. H. \& Wiesner, S. J., Communication via one-and two-particle operators on Einstein-Podolsky-Rosen states. \textit{Phys. Rev. Lett.} \textbf{69}, 2881 (1992).

\bibitem{mattle1996dense} Mattle, K., Weinfurter, H., Kwiat, P. G. \& Zeilinger, A., Dense coding in experimental quantum communication. \textit{Phys. Rev. Lett.} \textbf{76}, 4656 (1996).

\bibitem{greenberger1990bell} Greenberger, D. M., Horne, M. A., Shimony, A. \& Zeilinger, A., Bell’s theorem without inequalities. \textit{Am. J. Phys} \textbf{58}, 1131 (1990).

\bibitem{li2012classification} Li, X. \& Li, D., Classification of general n-qubit states under stochastic local operations \& classical communication in terms of the rank of coefficient matrix. \textit{Phys. Rev. Lett.} \textbf{108}, 180502 (2012).

\bibitem{dur2000three} D{\"u}r, W., Vidal, G. \& Cirac, J. I., Three qubits can be entangled in two inequivalent ways. \textit{Phys. Rev. A} \textbf{62}, 062314 (2000).

\bibitem{raussendorf2001one} Raussendorf, R. \& Briegel, H. J., A one-way quantum computer. \textit{Phys. Rev. Lett.} \textbf{86}, 5188 (2001).

\bibitem{moreno2016remote} Moreno, M. G. M., Cunha, M. M. \& Parisio, F., Remote preparation of W states from imperfect bipartite sources. \textit{Quantum Information Processing} \textbf{15}, 3869-3879 (2016).

\bibitem{vaidman1999methods} Vaidman, L. \& Yoran, N., Methods for reliable teleportation. \textit{Phys. Rev. A} \textbf{59}, 116 (1999).

\bibitem{lutkenhaus1999bell} L{\"u}tkenhaus, N., Calsamiglia, J. \& Suominen, K-A., Bell measurements for teleportation. \textit{Phys. Rev. A} \textbf{59}, 3295 (1999).

\bibitem{bose} Bose, S. \& Home, D., Generic entangling through quantum indistinguishability. \textit{Pramana} \textbf{59}, 229-233 (2002).

\bibitem{felinto2006conditional} Felinto, D., \textit{et al.}, Conditional control of the quantum states of remote atomic memories for quantum networking. \textit{Nature Physics} \textbf{2}, 844-848 (2006).

\bibitem{yuan2007synchronized} Yuan, Z-S., \textit{et al.}. Pan, Synchronized independent narrow-band single photons and efficient generation of photonic entanglement. \textit{Phys. Rev. Lett.} \textbf{98}, 180503 (2007).

\bibitem{makino2016synchronization} Makino, K., \textit{et al.}, Synchronization of optical photons for quantum information processing. \textit{Science Advances} \textbf{2}, (2016).

\bibitem{divochiy2008superconducting} Divochiy, A., \textit{et al}, Superconducting nanowire photon-number-resolving detector at telecommunication wavelengths. \textit{Nature Photonics} \textbf{2}, 302-306 (2008).

\bibitem{0953-2048-28-10-104001} Mattioli, F., \textit{et al.}, Photon-number-resolving superconducting nanowire detectors. \textit{Superconductor Science and Technology} \textbf{28}, 104001 (2015).

\bibitem{PhysRevLett.83.3103} White, A. G., James, D. F. V., Eberhard, P. H. \& Kwiat, P. G., Nonmaximally entangled states: Production, characterization, and utilization. \textit{Phys. Rev. Lett.} \textbf{83}, 3103 (1999).

\bibitem{xu2015experimental} Xu, F., \textit{et al.}, Experimental quantum key distribution with source flaws. \textit{Phys. Rev. A} \textbf{92}, 032305 (2015)

\bibitem{mizutani2015robustness} Mizutani, A., Imoto, N. \& Tamaki, K., Robustness of the round-robin differential-phase-shift quantum-key-distribution protocol against source flaws. \textit{Phys. Rev. A} \textbf{92}, 060303 (2015).

\bibitem{mendes2015femtosecond} Mendes, M. S., \textit{et al.}, Femtosecond source of unbalanced polarization-entangled photons. \textit{JOSA B} \textbf{32}, 1670-1675 (2015).

\bibitem{callen2006thermodynamics} Callen, H. B., {\em Thermodynamics \& an Intro. to Thermostatistics}
(John wiley \& sons, 2006).
	
\bibitem{pathria1972statistical} Pathria, R. K., {\em Statistical mechanics} (Elsevier, 1972).

\bibitem{bennett1996concentrating} Bennett, C. H., Bernstein, H. J., Popescu, S. \& Schumacher, Benjamin, Concentrating partial entanglement by local operations. \textit{Phys. Rev. A} \textbf{53}, 2046 (1996).

\bibitem{wootters1998entanglement} Wootters, W. K., Entanglement of formation of an arbitrary state of two qubits. \textit{Phys. Rev. Lett.} \textbf{80}, 2245 (1998).

\bibitem{coffman2000distributed} Coffman, V., Kundu, J. \& Wootters, W. K., Distributed entanglement. \textit{Phys. Rev. A} \textbf{61}, 052306 (2000).

\bibitem{fortescue2007random} Fortescue, B. \& Lo, H-K., Random bipartite entanglement from W and W-like states. \textit{Phys. Rev. Lett.}\textbf{98}, 260501 (2007).


\end{thebibliography}
\end{document}